\documentclass[10pt, preprint,aps, amssymb,amsmath,showkeys]{revtex4-1} 

\usepackage{amsthm}

\usepackage{amsmath}
\usepackage{amssymb}
\usepackage{graphicx}
\usepackage{epsfig}

\begin{document}
\author{Yamen Hamdouni}
\email{hamdouniyamen@gmail.com}
\affiliation{Department of physics, Faculty of Exact Sciences, Mentouri University, Constantine, Algeria}

\begin{abstract} We derive power series expansions for the magnetization, the internal energy, and the specific heat of the Heisenberg XX chain that are valid at low temperatures. The  coefficients of the series obtained  depend logarithmically on the fugacity. It is shown that depending on whether the magnetic field exceeds or not the critical point, the effects of either the coupling of the spins and the magnetic field can have different characters, as indicated by the different power laws established.   \end{abstract}
\keywords{Heisenberg XX chain; internal energy; magnetization; specific heat}
\title{The internal energy, the magnetization and the specific heat of the Heisenberg XX chain at low temperatures}

\maketitle
\section{Introduction}
The Heisenberg model plays an important role in the study of the magnetic properties of many materials \cite{kit}. Its success in providing ample explanations for  various phenomena occurring in   many-body spin systems made it a fundamental tool that has been extensively used  by many physicists. Historically, these investigations were motivated by the need to explore the properties of matter that arise from its periodic structure on the one hand, and  the effect of the accompanying quantum degrees of freedom of the individual atoms or molecules on the other hand. For instance, the Heisenberg model has been employed to investigate the low temperature properties of ferro and anti-ferromagnetic materials, where the notions of the quasi-particles called magnons and spinons  emerge, along with the corresponding transitions temperatures, i.e. the Curie and the N\'eel temperatues  that separate these phases from the paramagnetic one.\cite{phase} 

Furthermore,  the study of the  quantum phase transitions that occur at zero temperature, when a parameter of the system's Hamiltonian  crosses its critical value \cite{dev} have attracted great attention. The above processes turn out to be  quite relevant to a large number of applications, as is the case in the emerging field of quantum information technology. Indeed, in recent years, research in quantum theory has been widely oriented towards this new field. The interest in spin systems  stems from the fact that they are considered the most suitable candidates for the implementation of quantum computers, and quantum devices in a scalable manner \cite{loss}. 

The most notable difficulty when dealing with   many-body systems resides in the large number of the degrees of freedom characterizing them, which, in general, does not allow for a full analytical description of their properties. Spin lattices  are no exception to this  fact.
This is the reason for which numerical diagonalization techniques are invoked in many instances.
There exist, however, other techniques, like the powerful  Beth ansatz \cite{beth}, and  the Jordan-Wigner transformations \cite{Jordan} that may lead to exact results. It is obvious that such exact results are of great usefulness in obtaining a clear and plausible description of many-spin systems. Another important  fundamental result is the  Mermin--Wagner theorem \cite{mermin} which excludes any long-range order in low dimensional isotropic spin ferromagnets,  due to the thermal and quantum fluctuations. In other cases, it turns out that some simplifications and approximations may give rise to analytical results. A typical example is the long-wavelength approximation applied to the three-dimensional Heisenberg ferromagnet which yields the famous Bloch's law \cite{bloch},  valid only at low temperatures. Corrections to the latter law are obtained by taking into account magnon-magnon interactions \cite{dyson1, dyson2}, which Dyson called dynamical interactions between magnons. He also introduced what he named the kinematical interactions, which  arise from the finite dimensionality of the spin spaces of the atoms. His investigation results in a power series with respect to the temperature, which includes obviously the Bloch law as a special case.

The one-dimensional Heisenberg chain has been the subject of a large number of investigations~\cite{ lieb,taka,taka2,pfeuty,kat,baro,perk1,perk2,konto}. In particular, the $XY$ chain has been thoroughly studied by Katsura in \cite{kat}, and later generalized by Perk {\it et al}~\cite{perk1}. In most instances, the magnetization cannot be expressed in a simple mathematical form. The aim of this paper is to derive asymptotic expressions analogous to the Dyson ones  for the magnetization and the specific heat that are applicable to the Heisenberg XX chain at low temperatures. The paper is organized as follows: In section~\ref{sec2}, we introduce the model along with some mathematical preliminaries that are of interest for the subsequent discussion.  Section~\ref{sec3} deals with the case of a vanishing magnetic field. In Section~\label{sec4} we investigate the consequences of the presence of the critical point when the magnetic field is present  on the series expansions of the internal energy, magnetization and the specific heat. We end the paper with a brief discussion. 
\section{preliminaries \label{sec2}}

The one dimensional spin-$\frac{1}{2}$  Heisenberg $XX$ chain  is described by the Hamiltonian operator:
\begin{equation}
 \mathcal H=-J\sum\limits_{i=1}^N\Bigl( S_i^x S_{i+1}^x+ S_i^y S_{i+1}^y\Bigl)-h\sum_i^N S_i^z  \label{xy},
\end{equation}
where $\vec{\mathbf{S}}_i$ is the spin vector operator of the particle that is located at site $i$ of the  chain, and $h$ denotes the strength of the applied magnetic field which is pointing 
along the $z$ direction. In the case where $J$ is positive, the interaction between the spins is of ferromagnetic nature which will be assumed throughout the paper. On the contrary, when $J<0$, one is dealing with a antiferromagnetic model.
The Hamiltonian in equation~(\ref{xy}) can be exactly  diagonalized by using the Jordan-Wigner transformation \cite{Jordan}
\begin{eqnarray}
 S_j^+&=& c^\dag_j \exp\Bigl\{-i\pi \sum_{k=1}^{j-1}c_k^\dag c_k\Bigr\},\\
 S_j^-&=&  \exp\Bigl\{-i\pi \sum_{k=1}^{j-1}c_k^\dag c_k\Bigr\} c_j, \\
 S_j^z&=&c_j^\dag c_j-\frac{1}{2},
\end{eqnarray}
where $S_\pm=S_x\pm i S_y$, and  the $c_j$'s are fermionic operators satisfying $\{c_j, c^\dag_k\}=\delta_{jk}$, $\{c_j, c_k\}=\{c_j^\dag, c^\dag_k\}=0$.
In terms of the latter operators the Hamiltonian can be written as:
\begin{equation}
 \mathcal H=-J\sum_i(c_i^\dag c_{i+1}+c_{i+1}^\dag c_i)-h\sum_i(c_i^\dag c_i-\frac{1}{2}).
\end{equation}
Then by imposing the periodic boundary condition $c_{N+1}=c_1$, and by using the following Fourier transforms:
\begin{eqnarray}
 \eta_k=\frac{1}{\sqrt{N}}\sum_{j=1}^N e^{ijk}c_j, \\
 \eta_k^\dag=\frac{1}{\sqrt{N}}\sum_{j=1}^N e^{-ijk}c_j^\dag,
\end{eqnarray}
the resulting Hamiltonian becomes (we set $\hbar=1$)
\begin{equation}
 \mathcal H=\sum_k\omega_k \eta_k^\dag\eta_k+\frac{hN}{2}.
\end{equation}
The dispersion relation explicitly reads:
\begin{equation}
 \omega_k=-J\cos(k)-h,
\end{equation}
where $k=\frac{2\pi m}{N}$ with $m=-\frac{N}{2}, -\frac{N}{2}+1,\cdots \frac{N}{2}-1,\frac{N}{2}$ when $N$ is odd. For the case  of $N$ even, the possible values of $k$ are slightly different, since in this instance $m=-\frac{N-1}{2}, -\frac{N-2}{2},\cdots \frac{N-2}{2}, \frac{N-1}{2}$. Nevertheless, in the limit of $N$ very large ($N\to \infty$), the two cases become indistinguishable, in the sense that $k$ becomes a continuous variable whose domain determines the first Brillouin zone of the chain $-\pi\le k\le \pi$. The $XX$ spin chain has thus been mapped into a system of spinless free (non-interacting) fermions; it possesses a Fermi level only when $h<J$, for which the spectrum is gapless.

The internal energy  and the magnetization per site of the chain at temperature $T$ can be expressed in terms of the new operators as:
\begin{eqnarray}
 \frac{U}{N}&=&\frac{h}{2}-\frac{1}{N}\sum\limits_k \omega_k  \langle n_k \rangle\\
 M_z&=&-\frac{1}{2}+\frac{1}{N}\sum_k\langle n_k \rangle
\end{eqnarray}
where the mean number of the fermions in mode $k$ is given by the Fermi-Dirac distribution:
\begin{equation}
 \langle n_k \rangle=\frac{1}{e^{\beta\omega_k}+1}, \qquad n_k:=\eta_k^\dag\eta_k.
\end{equation}
  For convenience, we shall use, unless stated otherwise, the  inverse temperature $\beta=1/k_B T$ where $k_B$ is Boltzmann's constant.
  
Hence, in the continuum limit, we may write:
\begin{equation}
  \frac{U}{N}=\frac{h}{2}-J U_1-h U_2 \qquad  M_z=-\frac{1}{2}+U_2
\end{equation}
where
\begin{eqnarray}
 U_1&=&\frac{1}{2\pi}\int_{-\pi}^\pi \frac{ \cos(k) dk}{e^{-\beta(h+J\cos(k))}+1}, \\
 U_2 &=&\frac{1}{2\pi}\int_{-\pi}^\pi \frac{dk}{e^{-\beta(h+J\cos(k))}+1}.
\end{eqnarray}
The above  integrals cannot be carried out analytically, and one has to make recourse to numerical integration. As one may notice,  the dispersion relation of the model is exact and is valid for all values of the temperature, which means that the integrals should be carried out with respect to the full first Brillouin zone. Our aim is to derive series expansions analogous to the Dyson expansion, for the internal energy density, the specific heat and the magnetization per site  at low temperatures. 

For this purpose, let us first make the change of variable $\cos(k)=z$; then, we have
\begin{eqnarray}
U_1&=&\frac{1}{\pi}\int_{-1}^1 \frac{z dz}{\sqrt{1-z^2}} \frac{1}{e^{-\beta(h+J z)}+1},\\
 U_2&=&\frac{1}{\pi}\int_{-1}^1 \frac{dz}{\sqrt{1-z^2}} \frac{1}{e^{-\beta(h+J z)}+1}.
\end{eqnarray}
 Notice that the function $1/\pi\sqrt{1-z^2}$ is the density of states of the spin chain. The  points $z=\pm1$ where the above function diverges represent the Van Hove singularities of the chain. 

In the particular case when $h<J$, which corresponds to the Fermi level $k_F=\arccos(-h/J)$, we may further write
\begin{eqnarray}
U_1=\frac{1}{\pi}\int\limits_{-\frac{h}{J}}^1 \frac{z dz}{\sqrt{1-z^2}} \sum_{k=0}^\infty (-1)^k e^{-k\beta (h+J z)}-\frac{1}{\pi}\int\limits^{-\frac{h}{J}}_{-1} \frac{z dz}{\sqrt{1-z^2}} \sum_{k=1}^\infty (-1)^k e^{k \beta (h+J z)},\label{gap1} \\U_2= \frac{1}{\pi}\int\limits_{-\frac{h}{J}}^1 \frac{dz}{\sqrt{1-z^2}} \sum_{k=0}^\infty (-1)^k e^{-k\beta (h+J z)}-\frac{1}{\pi}\int\limits^{-\frac{h}{J}}_{-1} \frac{dz}{\sqrt{1-z^2}} \sum_{k=1}^\infty (-1)^k e^{k \beta (h+J z)} \label{gap2}.
\end{eqnarray}
Therefore, the lattice's ground state energy per site at zero temperature reads
 \begin{equation}
 \frac{U_0}{N}=-\frac{J}{\pi}\sqrt{1-\left(\frac{h}{J}\right)^2}-\frac{h}{\pi}\arcsin \left( \frac{h}{J} \right).
 \end{equation}
 Before we proceed further, for a reason that will become apparent shortly, we rewrite $U_1$ and $U_2$ in the convenient form:
  \begin{eqnarray}
 U_1 &= &\frac{1}{\pi}\sqrt{1- \left( \frac{h}{J} \right)^2}-\frac{2}{\pi}\sum_{k=1}^\infty (-1)^k  \cosh(k\beta h)\int^{0}_{-1}\frac{e^{k \beta J z}z  dz}{\sqrt{1-z^2}}+\mathcal G(h,J,\beta), \\
 U_2 &= &\frac{1}{2}+\frac{1}{\pi}\arcsin \left( \frac{h}{J} \right)-\frac{2}{\pi}\sum_{k=1}^\infty (-1)^k  \sinh(k\beta h)\int^{0}_{-1}\frac{e^{k \beta J z}z  dz}{\sqrt{1-z^2}}+\mathcal F(h,J,\beta),
\end{eqnarray}
where we have introduced the functions
\begin{eqnarray}
 \mathcal G(h,J,\beta)&=&-\frac{2}{\pi}\sum_{k=1}^\infty (-1)^k \cosh(k\beta h)\int^{\frac{h}{J}}_{0}\frac{e^{-k \beta J z} z dz}{\sqrt{1-z^2}}\nonumber\\&-& \frac{1}{\pi}\sum_{k=1}^\infty (-1)^k \int^{\frac{h}{J}}_{-\frac{h}{J}}\frac{e^{-k \beta (h-J z)} z dz}{\sqrt{1-z^2}},\\
  \mathcal F(h,J,\beta)&=&-\frac{2}{\pi}\sum_{k=1}^\infty (-1)^k \sinh(k\beta h)\int^{\frac{h}{J}}_{0}\frac{e^{-k \beta J z}  dz}{\sqrt{1-z^2}}\\&+& \frac{1}{\pi}\sum_{k=1}^\infty (-1)^k \int^{\frac{h}{J}}_{-\frac{h}{J}}\frac{e^{-k \beta (h-J z)}  dz}{\sqrt{1-z^2}}.
\end{eqnarray}

\section{ Chain in the absence of magnetic field  \label{sec3}}

The Fermi level in this case corresponds to the momenta

\begin{equation}
 k_F=\pm\pi/2.
\end{equation}
The functions $\mathcal F$ and $\mathcal G$ vanish for all values of $T$, meaning that $U_2=1/2$. Hence, the magnetization per site $M_z=-\frac{1}{2}+U_2$ is zero no matter what the temperature is, which means that there occurs no long-range order in the lattice.

The internal energy per site relative to the ground state  is given, on the other hand, by
\begin{eqnarray}
\frac{U-U_0}{N}&=&\frac{2J}{\pi}\sum_{k=1}^\infty (-1)^k  \int^{0}_{-1}\frac{e^{k \beta J z}z  dz}{\sqrt{1-z^2}}\nonumber \\
&=& J \sum_{k=1}^\infty (-1)^k \left[I_1(k\beta J)-L_{-1}(k\beta J)\right],
\end{eqnarray}
where $I_n(x)$ denotes the modified Bessel function of the first kind of order $n$, and $L_{n}(x)$ is the modified  Struve function of order $n$ 
 which satisfy~\cite{abra}
\begin{equation}
 L_n(z)-I_{-n}(z)\sim \frac{1}{\pi}\sum_{\nu=0}^\infty\frac{(-1)^{\nu+1} \Gamma(\nu+\frac{1}{2})}{\Gamma(-\nu+n+\frac{1}{2}) \left(\frac{z}{2}\right)^{2\nu-n+1}}, \label{struve}
\end{equation}
when $|z|$ is large. Hence, as $\beta\to\infty$:
\begin{eqnarray}
\frac{U-U_0}{N}&\sim& \frac{J}{\pi}\sum_{k=1}^\infty \left(\frac{1}{(J k \beta)^2}+\frac{21}{4(J k \beta)^4}+\frac{1395}{16(J k \beta)^6} +\cdots \right)\nonumber \\
&=& \frac{J}{\pi (J\beta)^2}\zeta(2)+ \frac{21J}{4\pi (J\beta)^4}\zeta(4)+ \frac{1395J}{16\pi (J\beta)^6}\zeta(6)+\cdots
\end{eqnarray}
where $\zeta(n)$ denotes the Riemann zeta function:
\begin{equation}
\zeta(n)=\sum_{k=1}^\infty \frac{1}{k^n}.
\end{equation}
Consequently, at low temperatures:
\begin{equation}
\frac{U-U_0}{N}\sim\left(\frac{\pi k_B^2}{6J}\right) T^2+\left(\frac{7\pi^3 k_B^4}{120 J^3}\right)  T^4+\left(\frac{31\pi^5 k_B^6}{336 J^5}\right)  T^6+\cdots
\end{equation}
From the latter expression of the internal energy, we deduce that the specific heat density of the lattice reads:
\begin{equation}
\frac {C}{k_BN}\sim\left(\frac{\pi k_B}{3J}\right) T+\left(\frac{7\pi^3 k_B^3}{30 J^3}\right)  T^3+\left(\frac{31\pi^5 k_B^5}{56 J^5}\right)  T^5+\cdots
\end{equation}
As a consequence, for sufficiently low temperatures,  the specific heat is nearly  linear with respect to the temperature.  By inspection, it turns out that expression~(\ref{asym3}) may be written  in the form  
\begin{equation} 
\frac {C}{k_B N}\sim a_1 T+a_3 T^3+a_5 T^5+\cdots,
\end{equation}
where the coefficients  in the latter series are:
\begin{equation}
a_{2n+1}=2(n+1)\frac{[(2 n + 1)!!]^2(\pi k_B)^{2n+1}}{J^{2 n + 1} (2n+1) }  (2 - 4^{-
     n }) \zeta(2 n + 2) .
\end{equation}
We thus conclude that when the magnetic field is absent, the internal energy at low temperatures is given by a series expansion that includes  only even powers of the temperature. The specific heat density  expands in a series that exhibits exclusively odd powers of the temperature.
\section{ Chain in a non-zero magnetic field\label{sec4}}
Let us begin by investigating the case $h<h_c=J$. The functions $\mathcal F$ and $\mathcal G$ introduced above do contribute to the internal energy and the magnetization. Actually,  for large values of  $\beta$ we can write:

\begin{eqnarray}
  U_2 &=& \frac{1}{2}+\frac{1}{\pi}\arcsin \left( \frac{h}{J} \right)-\frac{1}{\pi}\sum_{k=1}^\infty (-1)^k e^{k\beta h}\int^{0}_{-1}\frac{e^{k \beta J z} dz}{\sqrt{1-z^2}}+\mathcal F(h,J,\beta) \nonumber \\ &=& \frac{1}{2}+ \frac{1}{\pi}\arcsin \left( \frac{h}{J} \right)-\frac{1}{2}\sum_{k=1}^\infty (-1)^k e^{k\beta h} (I_0(\beta J k)-L_0(\beta J k))+\mathcal F(h,J,\beta). \label{noos2}
\end{eqnarray}
Using the asymptotic expression (\ref{struve}), we deduce that
\begin{small}\begin{eqnarray}
 U_2 &\sim &\frac{1}{2\pi}\sum_{k=1}^\infty (-1)^k e^{k\beta h}\left(-2\frac{\Gamma(\frac{1}{2})}{\Gamma(\frac{1}{2}) J \beta k}  + 2^3\frac{\Gamma(\frac{3}{2})}{\Gamma(-\frac{1}{2}) (J \beta k)^3}-2^5\frac{\Gamma(\frac{5}{2})}{\Gamma(-\frac{3}{2}) (J \beta k)^5}+\cdots\right)\nonumber \\
  &+& \frac{1}{2}+\frac{1}{\pi}\arcsin \left( \frac{h}{J} \right)+ \mathcal F(h,J,\beta).
\end{eqnarray}
\end{small}
Therefore:
\begin{eqnarray}
 U_2 &\sim & \frac{1}{2}+ \frac{1}{\pi}\arcsin \left( \frac{h}{J} \right)+\mathcal F(h,J,\beta)-\frac{1}{\pi\beta J}{\mathrm{ Li}}_{1}(-e^{\beta h})- \frac{1}{\pi(\beta J)^3}{\mathrm{ Li}}_{3}(-e^{\beta h})\nonumber \\ & -&  
 \frac{9}{\pi(\beta J)^5} \mathrm{ Li}_5(-e^{\beta h })-\cdots,
\end{eqnarray}
 where $\mathrm{ Li}_{n}$ denotes the Polylogarithmic function of order $n$. Taking into account the property \cite{wood}:
\begin{equation}
  \mathrm{ Li}_s(-e^{\mu })\sim -\frac{\mu^s}{\Gamma(s+1)}, \quad \mu\to\infty,\label{logasy}
\end{equation}
which implies that close to the absolute zero:
\begin{eqnarray}
  M_z &\to & \frac{1}{\pi}\arcsin \left( \frac{h}{J} \right)+\mathcal F(h,J,\beta)+\sum_{n=0}^\infty \frac{(h/J)^{2n+1} [(2n-1)!!]^2}{\pi (2n+1)!} \nonumber \\
  &=& \frac{2}{\pi}\arcsin \left( \frac{h}{J} \right)+\mathcal F(h,J,\beta),
  \end{eqnarray}
  where we can recognize the power series expansion of the $\arcsin$ function in the first line. 
  It immediately follows that $$\mathcal F(h,J,\beta)\sim -\frac{1}{\pi}\arcsin \left( \frac{h}{J} \right), \quad \beta\to\infty.$$

  Consequently,
 \begin{eqnarray}
  U_2&\sim & \frac{1}{2}-\frac{1}{\pi\beta J}{\mathrm{ Li}}_{1}(-e^{\beta h})- \frac{1}{\pi(\beta J)^3}{\mathrm{ Li}}_{3}(-e^{\beta h}) -  
 \frac{9}{\pi(\beta J)^5} \mathrm{ Li}_5(-e^{\beta h })-\cdots  \label{asym3}
\end{eqnarray}
In terms of the temperature, we may write for the magnetization: 
\begin{equation}
M_z \sim  -\mathcal Q_1(\mathfrak{z} ) T- \mathcal Q_{3}(\mathfrak{z}) T^3-\mathcal Q_{5}(\mathfrak{z}) T^5-\cdots \text{as} \quad  T\to 0, \label{bloch3}
\end{equation}
where $\mathfrak{z}$ is the fugacity 
\begin{equation}
 \mathfrak{z}=e^{\beta h},
\end{equation}
and in this case
\begin{equation}
 \mathcal Q_{2n+1}(\mathfrak{z}) =\frac{k_B^{2n+1} [(2n-1)!!]^2}{\pi J^{2n+1}}{\mathrm{ Li}}_{2n+1}(-\mathfrak{z}).
\end{equation}
We come to the  result that the magnetization expands in series of odd powers of the temperature, with coefficients that depend logarithmically on the fugacity of the lattice. 

In an analogous manner we find that:
\begin{eqnarray}
U_1\sim\frac{1}{\pi}\sqrt{1-\left(\frac{h}{J}\right)^2}+\mathcal G(h,J,\beta)+\mathcal R_{2}(\mathfrak{z}) T^2+\mathcal R_{4}(\mathfrak{z}) T^4+\mathcal R_{6}(\mathfrak{z}) T^6\cdots
\end{eqnarray}
where 
\begin{equation}
\mathcal R_{2n+2}(\mathfrak{z})=\frac{[(2n+1)!!]^2 k_B^{2n+2}}{\pi (2n+1) J^{2n+2}}{\mathrm{ Li}}_{2n+2}(-\mathfrak{z}).
\end{equation}
It can be shown  using equation (\ref{logasy}) that
\begin{equation}
\mathcal G(h,J,\beta)\sim \frac{1}{\pi}-\frac{1}{\pi}\sqrt{1-\left(\frac{h}{J}\right)^2}, \quad {\text {as}}\quad  T\to 0.
\end{equation}
Hence the internal energy per site behaves as:
\begin{eqnarray}
\frac{U}{N}&\sim&-\frac{J}{\pi}-J\left[\mathcal R_{2}(\mathfrak{z}) T^2+\mathcal R_{4} (\mathfrak{z})T^4+\mathcal R_{6}(\mathfrak{z}) T^6+\cdots\right]\nonumber\\ &+& h\left[\mathcal Q_{1}(\mathfrak{z}) T+\mathcal Q_{3}(\mathfrak{z}) T^3+\mathcal Q_{5}(\mathfrak{z}) T^5+\cdots\right].
\end{eqnarray}
This clearly shows that the interactions among the chain spins lead to a contribution to the internal energy with even powers with respect to the temperature. The contribution of the magnetic field, which is responsible of the magnetic order, leads however to odd powers of the temperature.

The quantities $\mathcal R$ and $\mathcal Q$ defined above satisfy the relations
\begin{eqnarray}
\frac{d\mathcal Q_{2n+1}(\mathfrak{z})}{dT}&=&-\frac{h(2n-1)}{J T^2}\mathcal R_{2n}(\mathfrak{z}),\\
\frac{d\mathcal R_{2n+2}(\mathfrak{z})}{dT}&=&-\frac{h(2n+1)}{J T^2}\mathcal Q_{2n+1}(\mathfrak{z}).
\end{eqnarray}
 This enables us to derive the asymptotic behavior of the specific heat at low temperatures, namely:
\begin{eqnarray}
\frac{C}{N}&\sim&\frac{\mathcal S_{-1}(\mathfrak{z})}{T}+\mathcal S_1(\mathfrak{z}) T+\mathcal S_3(\mathfrak{z}) T^3+\mathcal S_5(\mathfrak{z})T^5+\cdots\nonumber\\
&+&\mathcal P_{0}(\mathfrak{z})+\mathcal P_{2}(\mathfrak{z})T^2+\mathcal P_{4}(\mathfrak{z}) T^4+\cdots,
\end{eqnarray}
where:
\begin{eqnarray}
\mathcal S_{2n+1}(\mathfrak{z})&=&-[2(n+1)J+h^2(2n+1)/J]\mathcal R_{2n+2}(\mathfrak{z}),\\
\mathcal P_{2n}(\mathfrak{z})&=&2(2n+1)h\mathcal Q_{2n+1}(\mathfrak{z}).
\end{eqnarray}
It turns out that both the spin couplings and the magnetic field contribute to the specific heat  with odd powers. Even powers result solely from the magnetic field.

 When $h>h_c=J$ we find that 
\begin{eqnarray}
 U_2&=&\frac{1}{2\pi}\int_{-\pi}^\pi\sum_{k=0}^\infty (-1)^k e^{-\beta k(h+J \cos(z))}dz\nonumber\\
 &=& 1+\frac{1}{2\pi}\sum_{k=1}^\infty (-1)^k e^{-\beta k h}\int_{-\pi}^\pi e^{-k \beta J \cos(z)} dz\nonumber \\
 &=& 1+\sum_{k=1}^\infty (-e^{-\beta  h})^k I_0(k J \beta) .
\end{eqnarray}
Using the asymptotic expansion of the modified Bessel functions:
\begin{small}\begin{eqnarray} \displaystyle{ 
I_\nu(z) \sim  \frac{e^z}{\sqrt{2\pi z}}  \Bigl(1 - \frac{4 \nu^2 - 1}{8z} + \frac{(4 \nu^2 - 1) (4 \nu^2 - 9)}{2! (8z)^2} - \frac{(4 \nu^2 - 1) (4 \nu^2 - 9) (4 \nu^2 - 25)}{3! (8z)^3} + \cdots \Bigr)},
\end{eqnarray}\end{small}
for large $|z|$, we obtain
\begin{equation}\label{mess}
 U_2\sim1+\frac{1}{\sqrt{2\pi}}\sum_{k=1}^\infty \frac{(-e^{-\beta (h-J)})^k}{\sqrt{\beta J k}}\bigl(1+\frac{1}{8 J \beta}+ \frac{9}{2!(8 J \beta k)^2}+\cdots \bigr).
\end{equation}
That is
\begin{eqnarray}
 U_2&\sim&1+\frac{1}{\sqrt{2\pi\beta J}}{\mathrm{ Li}}_{\frac{1}{2}}(-e^{-\beta(h-J)})+ \frac{1}{8\sqrt{2\pi(\beta J)^3}}{\mathrm{ Li}}_{\frac{3}{2}}(-e^{-\beta(h-J)})\nonumber \\ &+&
 \frac{9}{2!\times 8^2\sqrt{2\pi(\beta J)^5}}{\mathrm{ Li}}_{\frac{5}{2}}(-e^{-\beta(h-J)})+\cdots \label{asym}
\end{eqnarray}

By inspection we can express the magnetization per spin in terms of  the temperature as:
\begin{equation}
M_z\sim\frac{1}{2}-\mathcal Q_{\frac{1}{2}}(\mathfrak{z}) T^{\frac{1}{2}}- \mathcal Q_{\frac{3}{2}}(\mathfrak{z}) T^{\frac{3}{2}}-\mathcal Q_{\frac{5}{2}}(\mathfrak{z}) T^{\frac{5}{2}}-\cdots \text{as} \quad  T\to 0, \label{bloch2}
\end{equation}
where:
\begin{equation}
 \mathcal Q_{\tfrac{2n+1}{2}}(\mathfrak{z}) =-\frac{k_B^{\tfrac{2n+1}{2}} [(2n-1)!!]^2}{n!\times 8^n\sqrt{2\pi J^{2n+1}}}{\mathrm{ Li}}_{\tfrac{2n+1}{2}}(-e^{\beta J}/\mathfrak{z}).
\end{equation}
We can evaluate $U_1$ in a similar way to  obtain:
\begin{equation}
 U_1\sim -\sum\limits_{n=0}^\infty \frac{k_B^{\frac{2n+1}{2}} \Gamma(n-\frac{1}{2}) \Gamma(n+\frac{3}{2})}{n! 2^n\sqrt{2\pi J^{2n+1}}} {\mathrm{ Li}}_{\tfrac{2n+1}{2}}(-e^{\beta J}/\mathfrak{z}).
\end{equation}

Therefore, the internal energy per site behaves at low temperatures as:
\begin{equation}
\frac{U}{N}\sim-\frac{h}{2}+\mathcal{R}_{\frac{1}{2}}(\mathfrak{z}) T^{\frac{1}{2}}+ \mathcal{R}_{\frac{3}{2}}(\mathfrak{z}) T^{\frac{3}{2}}+ \mathcal{R}_{\frac{5}{2}}(\mathfrak{z}) T^{\frac{5}{2}}+\cdots
\end{equation}
with
\begin{equation}
\mathcal{R}_{\frac{2n+1}{2}}(\mathfrak{z})=\left[h+\frac{J 2^{2n}\Gamma(n-\frac{1}{2}) \Gamma(n+\frac{3}{2})}{\pi  [(2n-1)!!]^2}\right] \mathcal Q_{\tfrac{2n+1}{2}}(\mathfrak{z}) =C_n  \mathcal Q_{\tfrac{2n+1}{2}}(\mathfrak{z}) .
\end{equation}
The situation  here is quite different in the sens that both the internal energy and the magnetization expand in series of half-integer powers of the temperature. That  means that the spins interactions along with the magnetic field ordering lead to the same asymptotic laws with respect to the temperature. 
After algebraic simplifications, we end up with the following asymptotic expansion of the specific heat per site:
\begin{equation}
\frac{C}{ N}\sim\frac{{\mathcal S}_{-\frac{3}{2}}(\mathfrak{z})}{T^{\frac{3}{2}}}+\frac{{\mathcal S}_{-\frac{1}{2}}(\mathfrak{z})}{T^{\frac{1}{2}}}+{\mathcal S}_{\frac{1}{2}}(\mathfrak{z})T^{\frac{1}{2}}+{\mathcal S}_{\frac{3}{2}}(\mathfrak{z})T^{\frac{3}{2}}+\cdots  \qquad \text{as} \ \  T\to 0,
\end{equation}
where 
\begin{eqnarray}
{\mathcal S}_{-\frac{3}{2}}(\mathfrak{z}) &=&\Bigl(\frac{h-J}{k_B}\Bigr)C_0 \sqrt{\frac{k_B}{J}}{\mathrm{ Li}}_{-\tfrac{1}{2}}(-e^{\beta J}/\mathfrak{z}),\\
{\mathcal S}_{\frac{2n-1}{2}}(\mathfrak{z})&=&\left[\frac{2n+1}{2}+\frac{(h-J)C_{n+1}}{J C_n }\left(\frac{4n^2+4n+1}{8(n+1)}\right)\right]\mathcal{R}_{\frac{2n+1}{2}}(\mathfrak{z}),\qquad n=0,1\cdots.
\end{eqnarray}
It can be seen that apart from the appearance of two  terms that exhibit negative half-integer powers, the specific heat displays the same power laws as the internal energy and the magnetization, regardless whether they originate from the coupling among the chain spins or from the effect of the applied magnetic field. Let us remark at the end that due to the fact that
\begin{equation}
 \lim\limits_{x\to 0^-} \frac{{\mathrm{ Li}}_{-\tfrac{n}{2}}(-e^{a/ x})}{x^b}=0
\end{equation}
for all positive $a$ and $b$, the specific heat effectively tends to zero at the absolute zero, as should be.
\section{Discussion and concluding remarks}
In this paper we have investigated  the low-temperature behavior of the magnetization, the internal energy, and the specific heat of the Heisenberg XX chain with and without the applied magnetic field. It turns out that when the magnetic field does not exceed the critical value, the coupling between the spins of the chain results in an expansion of the internal energy that displays only even powers with respect to the temperature, the coefficients of which depend logarithmically on the fugacity. The contribution of the magnetic field, when present, yields odd powers. The latter results may be accounted for by the differences in the effects the coupling and the magnetic field exert on the free fermions to which the spins of the chain are mapped; hence  they can provide a tool to distinguish the characteristics of either property of the chain at sufficiently low temperatures. The magnetization in this case, being barely a consequence of the magnetic field, expands in a series of odd powers. When the magnetic field exceeds the critical point, we find that all the above quantities exhibit expansions in half-integer powers of the temperature, which indicates that the coupling among the chain's spins as well as  the magnetic field  yield  nonlocal effects of essentially the same nature. The differences thus established are due to the presence of the critical point at which the lattice spectrum changes its nature from being  gapless.

\end{document}